\documentclass{article}
\usepackage{spconf,amsmath,graphicx}

\usepackage{multicol, multirow}
\usepackage{adjustbox}
\usepackage{amssymb}
\usepackage{mathtools}
\usepackage{makecell, booktabs}
\usepackage{url, cite}
\usepackage{tabularx}
\usepackage{hyperref}
\usepackage{xcolor}
\hypersetup{
    colorlinks=true,
    linkcolor=purple,
    filecolor=magenta,      
    urlcolor=purple,
}
\newcolumntype{Y}{>{\centering\arraybackslash}X}

\title{DcaseNet: An integrated pretrained deep neural network for detecting and classifying acoustic scenes and events}
\name{Jee-weon Jung$^{1,2}$, Hye-jin Shim${^1}$, Ju-ho Kim${^1}$, and Ha-Jin Yu${^1}$\sthanks{$^*$Corresponding author.}\thanks{This research was supported by Basic Science Research Program through the National Research Foundation of Korea(NRF) funded by the Ministry of Science, ICT \& Future Planning(2020R1A2C1007081).}}
\address{$^1$School of Computer Science, University of Seoul, $^2$Naver Corporation}
\begin{document}
\ninept
\maketitle

\begin{abstract}
Although acoustic scenes and events include many related tasks, their combined detection and classification have been scarcely investigated. 
We propose three architectures of deep neural networks that are integrated to simultaneously perform acoustic scene classification, audio tagging, and sound event detection. 
The first two architectures are inspired by human cognitive processes. 
The first architecture resembles the short-term perception for scene classification of adults, who can detect various sound events that are then used to identify the acoustic scene. 
The second architecture resembles the long-term learning of babies, being also the concept underlying self-supervised learning. Babies first observe the effects of abstract notions such as gravity and then learn specific tasks using such perceptions.  
The third architecture adds a few layers to the second one that solely perform a single task before its corresponding output layer. 
The aim is to build an integrated system that can serve as a pretrained model to perform the three abovementioned tasks. 
Experiments on three datasets demonstrate that the proposed architecture, called DcaseNet, can be either directly used for any of the tasks while providing suitable results or fine-tuned to improve the performance of one task. 
The code and pretrained DcaseNet weights are available at \url{https://github.com/Jungjee/DcaseNet}. 
\end{abstract}

\section{Introduction}
\label{sec:intro}
Recent advances in deep learning have led to improved performance of acoustic scene classification (ASC) and event related systems in various applications \cite{Koutini2019TheClassification, Mun2017GENERATIVEHYPER-PLANE, Chen2019IntegratingModeling, Barchiesi2015AcousticProduce, Bisot2016AcousticLearning, Jung2017DNN-basedDuplication}. 
The detection and classification of acoustic scenes and events (DCASE) community holds annual challenges with public datasets \cite{Mesaros2018AcousticEntries, Plumbley2018ProceedingsDCASE, Mandel2019ProceedingsDCASE}. 
The DCASE challenge datasets have fostered research on various tasks including ASC, audio tagging (TAG), sound event detection (SED), bird audio detection, and sound localization. 

However, these tasks have been independently studied using different deep neural networks (DNNs), despite that their characteristics and required information are highly related. 
Few studies have explored DNN architectures that use two tasks. 
Jung \textit{et al.} \cite{Jung2020AcousticTagging,Kim2020AUDIOCLASSIFICATION} applied an attention mechanism for ASC using a pretrained TAG system. 
The study was inspired by human cognition (e.g., \cite{Guastavino2007CategorizationProject}), as humans first perform SED and leverage this information to classify scenes. 
For instance, perceiving car horns and traffic sounds can be helpful for knowing that he/she is standing in a street. 
Imoto \textit{et al.} \cite{Imoto2020SoundLabels, Tonami2019JointLearning} explored the relation between ASC and SED, proposing DNNs to perform the two tasks simultaneously through a multi-task learning framework \cite{Caruana1997MultitaskLearning}. 
However, the integration of DNNs for related tasks remains in a preliminary stage because only pairs of two related tasks are investigated and motivations on the relationship of these tasks has not been explored. 
Hence, studies which address motivations for using multiple tasks and analyze the relationship between such tasks requires further investigation. 

\begin{figure*}
    \centering
    \includegraphics[width=\textwidth]{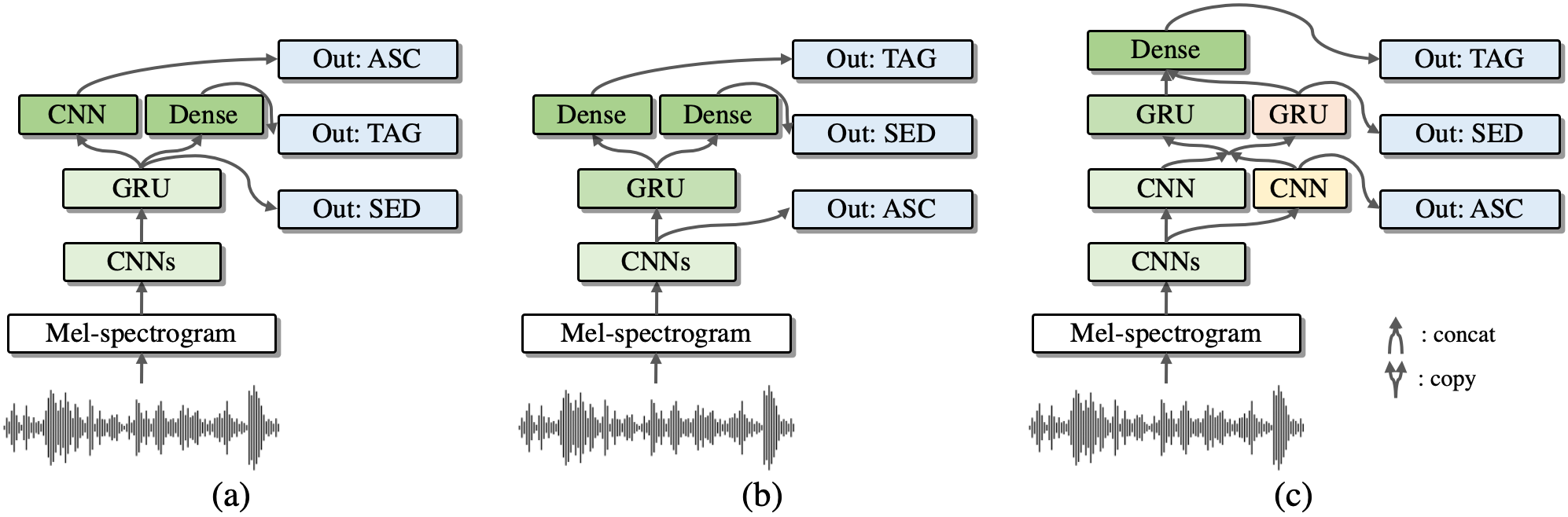}
    \caption{Proposed DNN architectures (DcaseNets). 
    (a) DNN inspired by an adult’s perception mechanism of acoustic scenes that uses event detection for scene classification \cite{Guastavino2007CategorizationProject}. 
    (b) Network performing ASC first considering the required abstraction level of representations, inspired by baby’s long-term learning procedure. 
    (c) Network extending that in (b) by adding a few separate layers before each task output. 
    The separate layers are concatenated with the mainstream information path (green boxes) before being feed-forwarded.}
    \label{fig:architectures}
\end{figure*}

We explore the integration of three tasks by proposing different DNN architectures (Fig. \ref{fig:architectures}), from which two are inspired by human cognitive processes and the third extends the second architecture. 
The integrated framework performs two segment-level tasks and one frame-level task using a single DNN that jointly learns and performs ASC, TAG, and SED (three tasks are introduced in details in Section \ref{sec:related}). 
The first architecture resembles the short-term perception of adults for ASC, being similar to the method used by Jung \textit{et al.} \cite{Jung2020AcousticTagging}. 
Instead of using a TAG system to improve an ASC system, we propose an integrated framework that first performs SED followed by ASC and TAG. 

The second architecture resembles the long-term learning of babies, which is also a motivation for self-supervised learning studies \cite{BengioAAAI2020ProgramCustom, Jing2020Self-supervisedSurvey, Doersch2015UnsupervisedPrediction}. 
In self-supervised learning, pretraining a DNN for relatively abstract tasks and then fine-tuning it for a specific task is an effective approach. 
It mimics babies first acquiring abstract notions (e.g., the effects of gravity) and then learning specific tasks based on the corresponding perceptions. 
Likewise, we assume that segment-wise multiclass ASC might require a relatively low abstraction level compared with SED, which requires frame-wise multilabel binary classification. 
Thus, in the proposed architecture, we consider the abstraction level of each task and perform relatively coarse scene classification followed by specific SED. 
Comparative experiments demonstrate that the second architecture provides high performance, and thus we extend it using a few separate layers for each task. 

The goal of this study is to build a single DNN that integrates ASC, TAG, and SED and can be fine-tuned to emphasize the performance of any of these tasks. 
The proposed DNNs, called DcaseNets, are intended to establish pretrained models for a wide range of acoustic tasks, like the ImageNet pretrained model \cite{Deng2009ImageNet:Database}. 
The main contributions of this study are threefold:
\begin{enumerate}
    \item Integrating DNN architectures, DcaseNets, to simultaneously perform ASC, TAG, and SED. 
    \item Developing DNN architectures based on human cognitive processes, namely, scene perception of adults and long-term learning of babies. 
    \item Demonstrating that fine-tuning the integrated DNNs for a specific task improves performance. 
\end{enumerate}

The remainder of this paper is organized as follows. 
In Section \ref{sec:related}, we describe ASC, TAG, and SED and discuss early studies on their integration. 
In Section \ref{sec:proposed}, we explain the proposed DNNs for setting integrated architectures to perform the three tasks. 
In Section \ref{sec:exp_res}, we report the experiments and results to verify the architecture performance. Finally, we draw conclusions in Section \ref{sec:conclusion}. 

\section{Related Work}
\label{sec:related}
ASC is a multiclass classification task that identifies a segment as being one of predefined scenes (i.e., classes). 
Acoustic scenes have an abstract (i.e., ambiguous) definition, and thus various characteristics may coincide across different scenes \cite{Jung2019DistillingClassification, Jung2020KnowledgeClassification}. 
For example, `airport' and `shopping\_mall', which are both predefined scenes in the DCASE ASC challenge, may contain people talking and the acoustic properties of large indoor spaces. 
These characteristics lead to a large intraclass variability, possibly degrading the classification performance. 
In addition, recent studies on ASC have addressed the impact of different recording devices on classification \cite{Primus2019ACOUSTICDEVICES, Gharib2018UNSUPERVISEDCLASSIFICATION, Kosmider2019CalibratingDevices}. 

TAG and SED are both multilabel binary classification tasks that perform event detection (i.e., judge the presence of various sounds). 
TAG conducts segment-level event detection and provides a vector in which each dimension is a real number between 0 and 1, indicating the probability of presence of a sound event throughout an input segment. 
SED conducts frame-level event detection and provides a matrix in which each row indicates the presence of sound events in a frame. 
Therefore, SED performs TAG with the onset and offset information from each sound event. Various studies on deep learning have addressed frame-level and the subsequent aggregation into segment-level classification. Likewise, we perform SED before TAG in the three proposed DNN architectures. 

Scarce research on the combination of related sound classification tasks has been conducted \cite{Imoto2020SoundLabels,Tonami2019JointLearning, Jung2020AcousticTagging, Kim2020AUDIOCLASSIFICATION}. 
Imoto \textit{et al.} \cite{Imoto2020SoundLabels} assumed that ASC and SED are related and performed them simultaneously using a multitask learning framework. 
Jung \textit{et al.} \cite{Jung2020AcousticTagging} aimed to mimic the human perception mechanism of leveraging TAG for ASC by applying an attention mechanism using the output of a pretrained TAG network. 
However, we assume that analysis and architecture designs accounting for the task relations can further improve integrated systems.

\section{Integrated Framework}
\label{sec:proposed}
We propose and investigate three DNN architectures to jointly perform ASC, TAG, and SED, as illustrated in Fig. \ref{fig:architectures}. 
The architectures are called as \textit{DcaseNet} because it jointly detects and classifies acoustic scenes and events. 
The design choice of the first and second architectures (Fig. \ref{fig:architectures}-(a) and -(b)) is inspired by human cognitive processes. 
The first architecture resembles the perception mechanism of an adult, expanding the method in \cite{Jung2020AcousticTagging}. 
This architecture first performs the event detection (i.e., SED) using a convolutional recurrent neural network (CRNN) to then perform segment-level scene classification (i.e., ASC). 
From the two event detection tasks, TAG and SED, we perform SED in lower layers (closer to the input layer) in the three proposed architectures by assuming that detecting short events improves the overall detection throughout an audio segment. 

The second architecture (Fig. \ref{fig:architectures}-(b)) resembles long-term learning of babies. 
First, abstract notions (e.g., gravity and dimension) are acquired to then perform specific tasks (e.g., moving things) using these notions. 
Studies on self-supervised learning \cite{BengioAAAI2020ProgramCustom, Jing2020Self-supervisedSurvey, Doersch2015UnsupervisedPrediction} are also based on this process to pretrain DNNs with unlabeled data. 
Similarly, we consider the required abstraction level in each task representation. 
We assume that ASC requires a relatively lower abstraction level than both TAG and SED. 
Hence, the second architecture performs ASC after a few convolutional neural network (CNN) blocks and then performs TAG and SED after a gated recurrent unit (GRU) layer. 

As the two architectures are inspired by human cognition, we experimentally test their performances and find that the second architecture tends to outperform the first one. 
Thus, the third architecture (Fig. \ref{fig:architectures}-(c)) extends the second architecture with additional layers before conducting each task. 
These separate layers are also concatenated and feed-forwarded to subsequent layers. 
As applied in \cite{He2016IdentityNetworks}, we aim to maintain an information path (green layers in Fig. \ref{fig:architectures}) while dedicating few layers to solely concentrate on performing an individual task. 
In addition, instead of performing TAG and SED in parallel, hidden layers perform SED and TAG in sequence, assuming that TAG, a segment-level task, requires a relatively higher abstraction level than SED, a frame-level task. 
For the three architectures, we first perform single and joint training. 
The models trained for single tasks establish the baselines for the corresponding architectures. 
After training, we fine-tune every model by using jointly trained DNNs for each task to determine the final performance on the target task. 

Each component of DcaseNet is based on high-performing architectures for each task. 
The CRNN (two light green boxes in Fig. \ref{fig:architectures}) in the three DcaseNet architectures is the one used by Cao \textit{et al.} \cite{Cao2019PolyphonicStrategy}, who achieved the second place in task 3 of the DCASE 2019 challenge using this network. 
The CRNN adopts eight convolutional layers with batch normalization \cite{Ioffe2015BatchShift}, followed by a bidirectional GRU layer. 
`CNN' at the top of Fig. \ref{fig:architectures}-(a) establishes a residual block and was implemented in \cite{Kim2020AUDIOCLASSIFICATION}, which was submitted for task 1-a in the DCASE 2020 challenge. 
`Dense' block in the three architectures before the TAG output is the one used by Akiyama \textit{et al.} \cite{Akiyama2019DCASETagging}, who won task 2 of the DCASE 2019 challenge. 
It adopts two fully-connected layers followed by dropout \cite{Srivastava2014Dropout:Overfitting} and ReLU non-linear activation. 
`Dense' block before the SED output in Fig. \ref{fig:architectures}-(b) also comprises two fully-connected layers with dropout and ReLU activation. 

\begin{table}[!t]
    \centering
    \begin{tabularx}{\linewidth}{@{}l|YYY@{}}
        \Xhline{1pt} 
        Task & ASC & TAG & SED\\
        \Xhline{1pt}
        Train duration (hours) & 43.0 & 10.5 & 10.0\\
        \# Train segments & 13,965 & 3,976& 600\\
        Segment duration (seconds) & 10 & 0.3$\sim$30& 60 \\
        \# Evaluation segments & 2,970 & 994 & 100\\
        \# Classes & 10 & 80 & 14\\
        \Xhline{1pt}
    \end{tabularx}
    \caption{Specifications of datasets for joint training of proposed DcaseNets. `ASC': Task 1-a of DCASE 2020 challenge, `TAG': Task 2 of DCASE 2019 challenge, `SED': Task 3 of DCASE 2020 challenge.}
    \label{tab:DBs}
\end{table}

\section{Experiments and results}
\label{sec:exp_res}

\subsection{Datasets and metrics}
\label{ssec:DB}
We use the DCASE 2020 Task 1-a dataset for the ASC Task, DCASE 2019 Task 2 dataset for the TAG task, and DCASE 2020 Task 3 dataset for the SED task. 
The dataset specifications including duration, number of segments, and number of classes are listed in Table \ref{tab:DBs}. 
For TAG, we only use the curated training set, thus excluding the noisy set and also split 20\% of the dataset to report evaluation performance, because the labels for the challenge evaluation set are not publicly available and a separate validation set does not exist\footnote{The train/test split we used is available along with our code in the Github repository.}.
We resampled all the segments to a rate of 24 kHz and used 16-bit resolution and monaural audio. 

The performances of the proposed architectures are reported using four metrics: overall classification accuracy (Acc) for ASC, label-weighted label-ranking precision (lwlrap) for TAG, and F-score (F1) and error rate (ER) for SED. 
Higher values indicate better results for all metrics except for ER. 
For brevity, we omit the detailed instructions on each metric, which can be found on the DCASE website.

\subsection{Experimental configurations}
\label{ssec:exp_config}
We use the 128-dimensional Mel-spectrograms as input features to each DNN. The spectrograms are extracted using 2,048 bins of the fast Fourier transform and 40 ms windows with 20 ms overlap. 
To train using multiple datasets concurrently, we configure 500 iterations as one epoch and train 160 epochs. 
The batch sizes for ASC, TAG, and SED tasks are 32, 24, and 32, respectively. 
During training for ASC and TAG, the segment duration is randomly cropped to 5 and 30 s for SED to construct the mini-batch and obtain data augmentation effect. 
We use Adam optimization \cite{Kingma2015Adam:Optimization} with a learning rate of 0.001. 
For both joint training and fine-tuning towards each task, the hyperparameters are not changed. 

The common CRNN for the three architectures comprises eight convolutional layers followed by batch normalization layers, where the last convolutional layer has 512 output filters. 
A bidirectional GRU layer with 512 nodes is used. 
`Dense' block before the TAG output comprises two fully-connected layers with 1,024 nodes per layer. 
Other detailed configurations can be found in the Github repository of this study.

\begin{table}[!t]
    \centering
    \begin{tabularx}{\linewidth}{@{}l|YYYY@{}}
         \Xhline{1pt}
         \multirow{2}{*}{Architecture} & ASC & TAG & \multicolumn{2}{c}{SED}\\
          & Acc & lwlrap & F1 & ER\\
         \Xhline{1pt}
         DCASE baseline & 54.10 & - & 60.60 & 0.5400\\
         Reference systems & 65.30 & - & 76.20 & 0.3050\\
         \hline
         DcaseNet-v1 (Fig. & \multirow{2}{*}{67.68} & \multirow{2}{*}{68.01} & \multirow{2}{*}{74.80} & \multirow{2}{*}{0.3486}\\
         \ref{fig:architectures}-(a))(w/o Mix-up)\\
         DcaseNet-v1 (Fig. & \multirow{2}{*}{68.19} & \multirow{2}{*}{69.41} & \multirow{2}{*}{\textbf{79.62}} & \multirow{2}{*}{\textbf{0.2926}}\\
         \ref{fig:architectures}-(a))(w/ Mix-up)\\
         DcaseNet-v2 & \multirow{2}{*}{\textbf{69.54}} & \multirow{2}{*}{69.19} & \multirow{2}{*}{79.34} & \multirow{2}{*}{0.3085}\\
         (Fig. \ref{fig:architectures}-(b))\\
         DcaseNet-v3 & \multirow{2}{*}{68.33} & \multirow{2}{*}{\textbf{70.62}} & \multirow{2}{*}{78.61} & \multirow{2}{*}{0.3085}\\
         (Fig. \ref{fig:architectures}-(c))\\
         \Xhline{1pt}
    \end{tabularx}
    \caption{Performance of proposed architectures for the performance for each task along with results from the DCASE official baseline and recent methods (higher values indicate better performance for all metrics except for ER). The performance of the reference methods are taken from \cite{Kim2020AUDIOCLASSIFICATION} for ASC and \cite{Nguyen2020DCASETRACKING} for SED. The values in boldface indicate the highest performance per metric.}
    \label{tab:baselines}
\end{table}
\begin{table*}[!t]
    \centering
    \begin{tabularx}{\linewidth}{@{}l|Y|YYY|YYYY|YYYY@{}}
         \Xhline{1pt}
         \multirow{3}{*}{Architecture} & \multirow{3}{*}{\#Param} & \multicolumn{3}{c}{Task combination} &  \multicolumn{4}{c}{Joint training (rand init)} & \multicolumn{4}{c}{Fine-tune (joint train init)}\\
         \cmidrule(lr){3-5}\cmidrule(lr){6-9}\cmidrule(lr){10-13}
         & & \multirow{2}{*}{ASC} & \multirow{2}{*}{TAG} & \multirow{2}{*}{SED} & ASC & TAG & \multicolumn{2}{c}{SED} & ASC & TAG & \multicolumn{2}{c}{SED}\\
          & & & & & Acc & lwlrap & F1 & ER & Acc & lwlrap & F1 & ER\\
         \Xhline{1pt}
         \multirow{2}{*}{DcaseNet-v1} & 8.7M&\checkmark & \checkmark & $\times$ & 59.23 & 53.55 & - & - & 67.18 & \textbf{68.81} & - & -\\ 
         \multirow{2}{*}{(Fig. \ref{fig:architectures}-(a))} & 8.7M&\checkmark & $\times$ & \checkmark & 57.04 & - & 64.75 & 0.4939 & 66.84 & - & \textbf{75.86} & \textbf{0.3365}\\ 
         \multirow{2}{*}{(w/o Mix-up)} & 8.7M&$\times$ & \checkmark & \checkmark & - & \textbf{71.91} & 74.66 & 0.3597 & - & \textbf{73.39} & \textbf{77.80} & \textbf{0.3153}\\ 
         & 8.7M&\checkmark & \checkmark & \checkmark & 56.13 & 59.25 & 59.82 & 0.5367 & 67.08 & \textbf{71.80} & \textbf{75.71} & \textbf{0.3470}\\
         \hline\hline
         \multirow{2}{*}{DcaseNet-v1} & 8.7M&\checkmark & \checkmark & $\times$ & 55.19 & 59.53 & - & - & 67.08 & \textbf{70.56} & - & -\\ 
         \multirow{2}{*}{(Fig. \ref{fig:architectures}-(a))} & 8.7M&\checkmark & $\times$ & \checkmark & 60.27 & - & 63.42 & 0.5097 & \textbf{68.73} & - & 78.31 & 0.3079\\ 
         \multirow{2}{*}{(w/ Mix-up)} & 8.7M&$\times$ & \checkmark & \checkmark & - & \textbf{75.18} & 75.27 & 0.3645 & - & \textbf{74.83} & 78.93 & 0.2963\\ 
         & 8.7M&\checkmark & \checkmark & \checkmark & 57.01 & 65.22 & 66.99 & 0.4764 & \textbf{68.59} & \textbf{71.64} & 78.06 & 0.3238\\
         \hline\hline
         & 8.9M&\checkmark & \checkmark & $\times$ & 58.92 & 60.00 & - & - & \textbf{69.87} & \textbf{69.68} & - & -\\ 
         DcaseNet-v2 & 8.9M&\checkmark & $\times$ & \checkmark & 58.79 & - & 69.53 & 0.4416 & 69.07 & - & 78.98 & \textbf{0.3058}\\ 
         (Fig. \ref{fig:architectures}-(b)) & 8.9M&$\times$ & \checkmark & \checkmark & - & \textbf{74.54} & 76.46 & 0.3571 & - & \textbf{74.76} & \textbf{81.32} & \textbf{0.2826}\\ 
         & 8.9M&\checkmark & \checkmark & \checkmark & 59.80 & 62.94 & 66.67 & 0.4817 & \textbf{69.57} & \textbf{71.38} & 79.26 & \textbf{0.2968}\\
         \hline\hline
         & 13.2M&\checkmark & \checkmark & $\times$ & 61.08 & 65.03 & - & - & \textbf{70.35} & \textbf{71.42} & - & -\\ 
         DcaseNet-v3 & 13.2M&\checkmark & $\times$ & \checkmark & 59.70 & - & 75.62 & 0.3512 & \textbf{69.44} & - & \textbf{79.61} & \textbf{0.2948}\\ 
         (Fig. \ref{fig:architectures}-(c)) & 13.2M&$\times$ & \checkmark & \checkmark & - & \textbf{76.23} & 77.93 & 0.3232 & - & \textbf{75.99} & \textbf{79.28} & \textbf{0.2958}\\ 
         & 13.2M&\checkmark & \checkmark & \checkmark & 56.80 & 70.40 & 75.19 & 0.3586 & \textbf{69.37} & \textbf{74.59} & \textbf{78.80} & 0.3185\\
         \Xhline{1pt}
    \end{tabularx}
    \caption{Experimental results on proposed integrated architectures (higher values indicate better performance for all metrics except for ER). 
    The baselines refer to single-task performance per architecture and correspond to those presented in Table \ref{tab:baselines}. 
    Column Fine-tune lists the results obtained from using the jointly trained model as pretrained network and performing fine-tune for each task. 
    The values in boldface indicate improved performance over the baseline.}
    \label{tab:exp_all}
\end{table*}

\subsection{Result analysis}
\label{ssec:result}
Table \ref{tab:baselines} lists the performance of the three DcaseNet architectures when trained on a single task. 
The top two rows describe the results from the official DCASE baselines and recent methods \cite{Kim2020AUDIOCLASSIFICATION, Nguyen2020DCASETRACKING}, both submitted for the DCASE 2020 challenge. 
Note that the reference performances for TAG cannot be reported because neither an official development nor an evaluation set exists. 
The proposed architectures show comparable performances with the recent methods, and hence their results are used as baselines in Table \ref{tab:exp_all}. 
In addition, the effect of Mix-up \cite{Zhang2018MixUp:Minimization} is investigated using the DcaseNet-v1 architecture. 
The results demonstrate that Mix-up is effective, as it improves the task performance in most cases. 
Thus, we applied Mix-up for both the DcaseNet-v2 and DcaseNet-v3 architectures. 

Table \ref{tab:exp_all} lists the performance after joint training and fine-tuning the three proposed architectures. 
For each architecture, the baselines are those listed in Table \ref{tab:baselines}. 
The task combinations for DNN training are indicated with checkmarks (e.g., the first row shows the results of the DNN trained using ASC and TAG). 
Column `Joint training' lists the results of training the model using different task combinations. 
Column `Fine-tune' lists the results of initializing the DNN using the jointly trained model and conducting fine-tuning for each task.

For DcaseNet-v1, which resembles an adult’s perception of scenes, the jointly trained model does not generalize well across the three tasks after fine-tuning. 
Joint training in the three tasks (fourth row) perform worse than the corresponding baseline in all four metrics. 
In addition, inconsistent results are obtained according to the application of Mix-up. 
When trained without Mix-up, the performance of TAG and SED improves, whereas with Mix-up, the performance of ASC and TAG improves. 
Even after fine-tuning, a single DNN that outperforms all three baselines cannot be obtained. 

DcaseNet-v2, which resembles a baby's learning procedure, outperforms DcaseNet-v1.  
After fine-tuning, DcaseNet-v2 shows higher performance than the corresponding baseline consistently, except for the model that jointly trained ASC and SED. 
Through a comparison between DcaseNet-v1 and DcaseNet-v2 architectures, we conclude that considering the abstraction level according to the assumed task complexity is more effective than mimicking an adult's perception mechanism. 
However, the jointly trained model (bottom row of DcaseNet-v2) has a lower overall performance than the corresponding baseline for each task without fine-tuning. 

DcaseNet-v3 achieves the highest performance among the three proposed architectures. 
The joint training model for the three tasks shows comparable performance to the corresponding baseline on each task. 
After fine-tuning for each task, DcaseNet-v3 outperforms the baseline except for the ER, which slightly increased from 0.3085 to 0.3185. 
In addition, joint training of ASC and TAG shows an accuracy of 70.35\% after fine-tuning for ASC. 
Overall, the results demonstrate the effectiveness of the architectures with few layers solely for each task. 
This outcome is consistent with the hidden layers being required to be fine-tuned when performing transfer learning, even though the pre-trained task has many similarities with the main task. 

Remarkably, for TAG, joint training with SED was the most effective approach across the three proposed architectures, possibly due to the close relation between TAG and SED. 
In addition, for joint training, degradation of accuracy in ASC requires further investigation.  
\vspace{-1pt}
\section{Conclusion and Future works}
\label{sec:conclusion}
We propose three integrated DNN architectures, DcaseNets, inspired by human cognitive processes to simultaneously perform ASC, TAG, and SED. 
The first architecture resembles the perception mechanism for acoustic scenes of adults, and the second architecture resembles the learning of babies, with the latter providing higher performance. 
The third architecture that adds a few layers before the output of each intermediate task further improves the performance of the second architecture. 
Jointly trained models can be further fine-tuned for each task. 
Experimental results show the high performance of the proposed DcaseNet-v3 for the three tasks after joint training, outperforming all the corresponding baselines with fine-tuning. 

However, as this is the first study that integrates the three related tasks, further investigation and improvements should be performed. 
As we experimentally verified that DcaseNet-v2 outperforms DcaseNet-v1, we will adopt self-supervised learning aiming to improve the task performance. 
In addition, we will apply knowledge distillation and leverage soft-label training. 
Soft-labels will enable cross-dataset training by, for example, generating soft-labels for the ASC dataset and using the result to calculate the TAG loss. 
\bibliographystyle{IEEEbib}
\bibliography{refs_mendeley, strings}

\begin{thebibliography}{10}

\bibitem{Koutini2019TheClassification}
K. Koutini, H. Eghbal-Zadeh, M. Dorfer, and G. Widmer,
\newblock ``{The receptive field as a regularizer in deep convolutional neural
  networks for acoustic scene classification},''
\newblock in {\em European Signal Processing Conference (EUSIPCO)}. 2019, IEEE.

\bibitem{Mun2017GENERATIVEHYPER-PLANE}
S. Mun, S. Park, D. Han, and H. Ko,
\newblock ``{Generative Adversarial Network Based Acoustic Scene Training Set
  Augmentation and Selection Using SVM Hyper-Plane},''
\newblock in {\em Proceedings of the Detection and Classification of Acoustic
  Scenes and Events Workshop (DCASE)}, 2017.

\bibitem{Chen2019IntegratingModeling}
H. Chen, Z. Liu, Z. Liu, P. Zhang, and Y. Yan,
\newblock ``{Integrating the Data Augmentation Scheme with Various Classifiers
  for Acoustic Scene Modeling},''
\newblock Tech. {R}ep., DCASE2019 Challenge, 2019.

\bibitem{Barchiesi2015AcousticProduce}
D. Barchiesi, D.~D. Giannoulis, D. Stowell, and M.~D. Plumbley,
\newblock ``{Acoustic Scene Classification: Classifying environments from the
  sounds they produce},''
\newblock {\em IEEE Signal Processing Magazine}, vol. 32, no. 3, pp. 16--34, 5
  2015.

\bibitem{Bisot2016AcousticLearning}
V. Bisot, R. Serizel, S. Essid, and G. Richard,
\newblock ``{Acoustic scene classification with matrix factorization for
  unsupervised feature learning},''
\newblock in {\em International Conference on Acoustics, Speech and Signal
  Processing (ICASSP)}. 2016, pp. 6445--6449, IEEE.

\bibitem{Jung2017DNN-basedDuplication}
J.-w. Jung, H.-S. Heo, I.-H. Yang, S.-H. Yoon, H.-j. Shim, and H.-J. Yu,
\newblock ``{DNN-based audio scene classification for DCASE 2017: dual input
  features, balancing cost, and stochastic data duplication},''
\newblock in {\em Proceedings of the Detection and Classification of Acoustic
  Scenes and Events Workshop (DCASE)}, 2017.

\bibitem{Mesaros2018AcousticEntries}
A. Mesaros, T. Heittola, and T. Virtanen,
\newblock ``{Acoustic scene classification: An overview of dcase 2017 challenge
  entries},''
\newblock in {\em International Workshop on Acoustic Signal Enhancement
  (IWAENC)}. 11 2018, pp. 411--415, IEEE.

\bibitem{Plumbley2018ProceedingsDCASE}
M.~D. Plumbley, C. Kroos, J.~P. Bello, G. Richard, and D.~P. Ellis,
\newblock {\em {Proceedings of the Detection and Classification of Acoustic
  Scenes and Events 2018 Workshop (DCASE)}},
\newblock Tampere University of Technology, 2018.

\bibitem{Mandel2019ProceedingsDCASE}
M. Mandel, J. Salamon, and D.~P. Ellis,
\newblock {\em {Proceedings of the Detection and Classification of Acoustic
  Scenes and Events 2019 Workshop (DCASE)}},
\newblock New York University, 2019.

\bibitem{Jung2020AcousticTagging}
J.-w. Jung, H.-j. Shim, J.-h. Kim, S.-b. Kim, and H.-J. Yu,
\newblock ``{Acoustic Scene Classification using Audio Tagging},''
\newblock in {\em Annual Conference of the ISCA, INTERSPEECH}, 2020.

\bibitem{Kim2020AUDIOCLASSIFICATION}
J.-h. Kim, J.-w. Jung, H.-j. Shim, and H.-J. Yu,
\newblock ``{Audio Tag Representation Guided Dual Attention Network for
  Acoustic Scene Classification},''
\newblock in {\em Proceedings of the Detection and Classification of Acoustic
  Scenes and Events Workshop (DCASE)}, 2020.

\bibitem{Guastavino2007CategorizationProject}
C. Guastavino,
\newblock ``{Categorization of environmental sounds Musikiosk View project
  Exploring sensory experience: Meaning and the senses View project},''
\newblock {\em Canadian Journal of Experimental Psychology}, vol. 61, no. 1,
  pp. 54, 2007.

\bibitem{Imoto2020SoundLabels}
K. Imoto, N. Tonami, Y. Koizumi, M. Yasuda, R. Yamanishi, and Y. Yamashita,
\newblock ``{Sound Event Detection by Multitask Learning of Sound Events and
  Scenes with Soft Scene Labels},''
\newblock in {\em International Conference on Acoustics, Speech and Signal
  Processing (ICASSP)}. 2020, pp. 621--625, IEEE.

\bibitem{Tonami2019JointLearning}
N. Tonami, K. Imoto, M. Niitsuma, R. Yamanishi, and Y. Yamashita,
\newblock ``{Joint analysis of acoustic events and scenes based on multitask
  learning},''
\newblock in {\em IEEE Workshop on Applications of Signal Processing to Audio
  and Acoustics}. 10 2019, vol. 2019-Octob, pp. 338--342, IEEE.

\bibitem{Caruana1997MultitaskLearning}
R. Caruana,
\newblock ``{Multitask Learning},''
\newblock {\em Machine Learning}, vol. 28, pp. 41--75, 1997.

\bibitem{BengioAAAI2020ProgramCustom}
Y. Bengio, G.~E. Hinton, and Y. LeCun,
\newblock ``Aaai2020 invited speaker program,'' 2020,
  \textit{https://aaai.org/Conferences/AAAI-20/invited-speakers/}.

\bibitem{Jing2020Self-supervisedSurvey}
L. Jing and Y. Tian,
\newblock ``{Self-supervised Visual Feature Learning with Deep Neural Networks:
  A Survey},''
\newblock {\em IEEE Transactions on Pattern Analysis and Machine Intelligence},
  pp. 1--22, 5 2020.

\bibitem{Doersch2015UnsupervisedPrediction}
C. Doersch, A. Gupta, and A.~A. Efros,
\newblock ``{Unsupervised visual representation learning by context
  prediction},''
\newblock in {\em IEEE/CVF International Conference on Computer Vision (ICCV)},
  2015, pp. 1422--1430.

\bibitem{Deng2009ImageNet:Database}
J. Deng, W. Dong, R. Socher, L.-J. Li, {Kai Li}, and {Li Fei-Fei},
\newblock ``{ImageNet: A large-scale hierarchical image database},''
\newblock in {\em IEEE/CVF Conference on Computer Vision and Pattern
  Recognition (CVPR)}. 3 2009, pp. 248--255, IEEE.

\bibitem{Jung2019DistillingClassification}
J.-w. Jung, H.-S. Heo, H.-j. Shim, and H.-J. Yu,
\newblock ``{Distilling the Knowledge of Specialist Deep Neural Networks in
  Acoustic Scene Classification},''
\newblock in {\em Proceedings of the Detection and Classification of Acoustic
  Scenes and Events Workshop (DCASE)}, 2019, pp. 114--118.

\bibitem{Jung2020KnowledgeClassification}
J.-w. Jung, H.-S. Heo, H.-j. Shim, and H.-J. Yu,
\newblock ``{Knowledge Distillation in Acoustic Scene Classification},''
\newblock {\em IEEE Access}, vol. 8, pp. 166870--166879, 9 2020.

\bibitem{Primus2019ACOUSTICDEVICES}
P. Primus and D. Eitelsebner,
\newblock ``{Acoustic Scene Classification with Mismatched Recording
  Devices},''
\newblock Tech. {R}ep., DCASE2019 Challenge, 2019.

\bibitem{Gharib2018UNSUPERVISEDCLASSIFICATION}
S. Gharib, K. Drossos, E. Cakir, D. Serdyuk, and T. Virtanen,
\newblock ``{Unsupervised Adversarial Domain Adaptation for Acoustic Scene
  Classification},''
\newblock in {\em Proceedings of the Detection and Classification of Acoustic
  Scenes and Events Workshop (DCASE)}, 2018.

\bibitem{Kosmider2019CalibratingDevices}
M. Kosmider,
\newblock ``{Calibrating neural networks for secondary recording devices},''
\newblock Tech. {R}ep., DCASE2019 Challenge, 2019.

\bibitem{He2016IdentityNetworks}
K. He, X. Zhang, S. Ren, and J. Sun,
\newblock ``{Identity mappings in deep residual networks},''
\newblock in {\em Lecture Notes in Computer Science}. 2016, vol. 9908 LNCS, pp.
  630--645, Springer Verlag.

\bibitem{Cao2019PolyphonicStrategy}
Y. Cao, Q. Kong, T. Iqbal, F. An, W. Wang, and M.~D. Plumbley,
\newblock ``{Polyphonic Sound Event Detection and Localization using a
  Two-Stage Strategy},''
\newblock in {\em Proceedings of the Detection and Classification of Acoustic
  Scenes and Events Workshop (DCASE)}. 2019, pp. 30--34, New York University.

\bibitem{Ioffe2015BatchShift}
S. Ioffe and C. Szegedy,
\newblock ``{Batch normalization: Accelerating deep network training by
  reducing internal covariate shift},''
\newblock in {\em International Conference on Machine Learning (ICML)}. 2015,
  vol.~1, pp. 448--456, IMLS.

\bibitem{Akiyama2019DCASETagging}
O. Akiyama and J. Sato,
\newblock ``{DCASE 2019 Task 2: Multitask Learning, Semi-supervised Learning
  and Model Ensemble with Noisy Data for Audio Tagging},''
\newblock in {\em Proceedings of the Detection and Classification of Acoustic
  Scenes and Events Workshop (DCASE)}, 2019, pp. 25--29.

\bibitem{Srivastava2014Dropout:Overfitting}
N. Srivastava, G. Hinton, A. Krizhevsky, I. Sutskever, and R. Salakhutdinov,
\newblock ``{Dropout: A simple way to prevent neural networks from
  overfitting},''
\newblock {\em Journal of Machine Learning Research}, vol. 15, pp. 1929--1958,
  2014.

\bibitem{Kingma2015Adam:Optimization}
D.~P. Kingma and J.~L. Ba,
\newblock ``{Adam: A method for stochastic optimization},''
\newblock in {\em International Conference on Learning Representations (ICLR)},
  2015.

\bibitem{Nguyen2020DCASETRACKING}
T.~N.~T. Nguyen, D.~L. Jones, and G.~W. Seng,
\newblock ``{DCASE 2020 Task 3: Ensemble of Sequence Matching Networks for
  Dynamic Sound Event Localization, Detection, and Tracking},''
\newblock Tech. {R}ep., DCASE2020 Challenge, 2020.

\bibitem{Zhang2018MixUp:Minimization}
H. Zhang, M. Cisse, Y.~N. Dauphin, and D. Lopez-Paz,
\newblock ``{MixUp: Beyond empirical risk minimization},''
\newblock in {\em International Conference on Learning Representations (ICLR)},
  2018.

\end{thebibliography}
\end{document}